# Liquid structure and temperature invariance of sound velocity in supercooled Bi melt

M. Emuna,[1] M. Mayo,[1] Y. Greenberg,[2] E. N. Caspi,[2] B.Beuneu,[3] E. Yahel[2] and G. Makov[1]

[1] *Materials Engineering Department, Ben-Gurion University of the Negev, 84105 Beer-Sheva, Israel.*

[2] *Physics Department, Nuclear Research Centre-Negev, 84190 Beer-Sheva, Israel.*

[3] *Laboratoire Léon Brillouin CEA/Saclay, 91191 Gif-Sur-Yvette Cedex, France.*

Structural rearrangement of liquid Bi in the vicinity of the melting point has been proposed due to the unique temperature invariant sound velocity observed above the melting temperature, the low symmetry of Bi in the solid phase and the necessity of overheating to achieve supercooling. The existence of this structural rearrangement is examined by measurements on supercooled Bi. The sound velocity of liquid Bi was measured into the supercooled region to high accuracy and it was found to be invariant over a temperature range of ca. 60 degrees, from 35 degrees above the melting point to ca. 25 degrees into the supercooled region. The structural origin of this phenomenon was explored by neutron diffraction structural measurements in the supercooled temperature range. These measurements indicate a continuous modification of the short range order in the melt. The structure of the liquid is analyzed within a quasi-crystalline model and is found to evolve continuously, similar to other known liquid pnictide systems. The results are discussed in the context of two competing hypotheses proposed to explain properties of liquid Bi near the melting: (i) liquid bismuth undergoes a structural rearrangement slightly above melting (ii) liquid Bi exhibits a broad maximum in the sound velocity located incidentally at the melting temperature.

## I. INTRODUCTION

The existence and nature of polyamorphism in elemental metals is the subject of intensive research.[1,2,3,4,5,6] Bi is a prime candidate elemental liquid for observation of liquid polyamorphism due to the semi-metallic electronic structure, which determines a low symmetry crystalline structure with multiple polymorphs, expressed in a complex pressure-temperature phase diagram.[7] Liquid Bi also exhibits a complex structure[8,9] and indeed, there have been several reports of liquid transformations in Bi. Notably a structural rearrangement identified at approximately 740°C[8] and associated with a change in configurational entropy.[10] Additional phase lines were identified to form a high pressure-high temperature liquid phase diagram in Ref. 11 on the basis of electrical resistivity measurements and these were later supported by high pressure diffraction studies along the melting curve.[12]

In addition, for almost 100 years there have been multiple indications of structural rearrangement in the liquid associated with the solidification of Bi. Apparently, attention to the problem was first drawn by G.I. Taylor on the basis of the cracked nature of elemental Bi crystals, leading P. Kapitza to suggest in 1928 the possible existence of an intermediate phase between the solid and the liquid.[13] Subsequent thermophysical studies of the solidification process did not find evidence in support for such a phase.[14]

However, a series of diffraction studies identified a significant change in the liquid structure as Bi was cooled into the supercooled regime.[15,16,17] The authors suggested that a significant, albeit continuous, rearrangement in the liquid structure occurs in the vicinity of the melting point.[16,17] Unfortunately, the quality of these structural diffraction studies was insufficient for a quantitative analysis of the changes in the liquid structure.[16]

Additional support for the hypothesis of structural rearrangement is provided by thermophysical studies. An experimental study found a strong correlation between the extent of overheating of liquid Bi above the melting temperature and the extent of subsequent supercooling achievable in contrast to normal liquid metals.[18] This indicates that upon melting, the liquid retains short order characteristics of its solid phase and additional thermal energy is required to attain the disordered liquid state which can then be supercooled.

Independently, Gitis and co-workers suggested the existence of a structural rearrangement above the melting point on the basis of the results of sound velocity measurements. In these studies[19,20,21], it was found that Bi, uniquely among the elements[22], exhibits a temperature invariant sound velocity over a range of ca. 30 degrees above melting, a result which was recently confirmed.[23] It was hypothesized[20] that this temperature invariance was a reflection of an underlying structural rearrangement in the liquid. Finally, we note that there have been additional suggestions that liquid Bi undergoes a structural rearrangement based on microstructural properties of resolidified Bi, e.g. Refs. 24,25,26.

Recently, we have performed detailed measurements of the temperature dependence of the structures of liquid Bi and Sb[8,27]. Advanced experimental and post-processing techniques allow us to achieve very high accuracy, both in the measured structure factor and in the radial distribution function (RDF) obtained from it.[28,29] Direct analysis of these results and further analysis within the quasi-crystalline model (QCM)[30], have not yielded any indication of significant changes in the liquid structure of Bi above the melting point.

In addition, very recently, we have studied the temperature dependence of the sound velocity in the isomorphic Bi-Sb alloy system[31] Sb is well known to exhibit a broad maximum in the sound velocity at approximately 895°C,[20,23] We found that all the alloy compositions studied exhibit similar broad maxima in the temperature dependence of their sound velocity, with the maximum temperature shifting down towards the liquidus temperature as the Bi fraction increases. This leads us to speculate that the plateau observed in the sound velocity of Bi may be typical of liquid pnictides and extend into a maximum in the supercooled regime with the anomaly being entirely coincidental. This may be consistent with the fact that no significant changes been identified in the structure of liquid Sb in the vicinity of the temperature of the sound velocity maximum. Instead, a continuous and monotonous evolution of the liquid structure is found throughout the measurement range of several hundred degrees, with similar structural processes occurring in the two closely related elements Sb and Bi.[27] These results do not support the existence of a significant structural rearrangement near the melting point, but, of course, do not rule out the possibility of such a structural rearrangement occurring in the liquid just below the melting point.

Therefore, the present study aims to resolve the century long question of the existence of a structural rearrangement in liquid Bi in the vicinity of the melting point by experimental studies in the supercooled region. Bi is able to undergo relatively extensive supercooling for an elemental liquid.[15,16,17] This raises the possibility that measurements of its thermophysical properties and structure in the supercooled region may shed light on structural rearrangements which may take place in liquid Bi near the melting point. No reports, that we are aware of, exist in the literature of sound velocity measurements in the supercooled state. In the real space, liquid Bi exhibits sub-structure that is apparent as a shoulder on the right hand side (r.h.s.) of the first peak in the radial distribution function. In an earlier study,[27] it has been shown that, for liquid Bi, the shoulder in the RDF decreases in magnitude with increasing temperature. Other studies have indicated that the shoulder may evolve into a peak in the supercooled region.[15,16] However, the quality of the experimental results was insufficient for a detailed analysis indicating small changes in the liquid's structure.[16] The metastable nature of the supercooled state requires specific experimental techniques, whereas the identification of any structural rearrangement requires highly accurate measurements, thus making this a challenging objective. In particular, we report on the first measurements of the sound velocity of supercooled liquid Bi over a range of ca. 25 degrees below the melting point. Using the methods developed in our studies of liquid pnictides, we obtain the structure of liquid Bi to high accuracy from several tens of degrees above melting to ca. 25 degrees below melting by neutron diffraction. These results are analyzed within the QCM to identify any changes in the liquid structure. Our findings are discussed in the context of the question of the existence of a structural rearrangement and in the context of the physics of liquid pnictides.

## II. EXPERIMENTAL

### A. Sound velocity of liquid Bi in the stable and supercooled region

The sound velocity of Bi above melting point was measured by the pulse-echo technique and the experimental setup has been described elsewhere.[31] The velocity of acoustic waves is calculated from the time interval of the transmitted wave, travelling a known distance in the melt. To achieve the accuracy needed, a modified pulse-echo technique was used to increase the accuracy by measuring the sound velocity differentially. The differential method involves frequent displacing of a buffer rod immersed in the melt, hence introducing perturbations in the liquid. Therefore, this method cannot be applied in the supercooled region, since the liquid is at a metastable state and will solidify promptly.

Alternatively, in the supercooled region, sound velocity was measured by a relative method, which the travelling time of the acoustic wave in the sample ($\Delta t_2$) in the supercooled state was compared to the travelling time measured near the melting point ($\Delta t_1$) and for which, the sound velocity $c_1$ was obtained using the differential method. The sound velocity ($c_2$) in the supercooled state was calculated from the ratio $c_2=c_1(\Delta t_1/\Delta t_2)$, where $c_1$ is the sound velocity at the melting point. During the measurements in the supercooled region, the buffer rod was positioned at a fixed distance from the bottom of the crucible.

The errors in determining the sound velocity originate from errors in determining the distance and the time interval required for the sound waves to travel through the sample. The nominal acoustic path was 1mm and the uncertainty in the acoustic path, resulting from the accuracy of the linear motor, is 1μm. The oscilloscope time base resolution for differential measurements was 2ns with a typical travel time of ca. 1μs. Thus, the total uncertainty in determining sound velocity above liquidus temperature is ca. 0.2%. Errors resulting from thermal expansions of the structural materials are estimated to be negligible in the modified pulse-echo method. The experimental uncertainties of the sound velocity, measured in the supercooled region, are similar, whereas we assume that due to the small temperature range, the thermal expansion of the components is negligible.

**B.    Neutron diffraction measurements**

Neutron diffraction measurements of liquid Bi were undertaken at the 7C2, a two-axis diffractometer at Laboratoire Léon Brillouin on the hot source of the Orphée reactor at Saclay, France. A monochromatic neutron beam, $\lambda=0.717Å$ was used with available momentum transfer range of 0.2–15.3Å$^{-1}$, at intervals of approximately 0.014Å$^{-1}$. Sample dimensions were 9.2mm in diameter and 60mm in length. Beam spot size was 10mm wide and 41mm high. The sample was mounted in a sealed vanadium capsule and the temperature was measured using thermocouples which were in contact with the bottom of the capsule. The use of a vanadium sample holder instead of silica, used in our previous studies,[8,27,29] enabled us to obtain a better signal to noise ratio, since a silica sample holder contributes 30% of the raw scattered intensity, in contrast to the vanadium sample holder which contributes less than 5% to total scattered intensity. The measurements were taken during intervals of 6-8 hours to obtain sufficiently good statistics, and also ensuring thermal equilibration of the system. Moreover, the sample cell was mounted in a cylindrical thin foil vanadium furnace, significantly longer than the sample's height. Such a furnace geometry minimizes temperature gradients along the sample. Data correction and structure factor calculations were performed using a program developed at the Laboratoire Léon Brillouin, whereas radial distribution functions, *g(r)*, were calculated as described in Ref. 28.

## III. RESULTS AND DISCUSSION

### A. The temperature dependence of sound velocity in liquid Bi and in the supercooled region

The sound velocity in liquid Bi was measured upon cooling through the temperature range from 803K to 520K. To validate the results, at least two data sets from the same sample were measured. The typical deviation between the different data sets was within the experimental errors. Above the melting temperature, measurement was carried out by the differential method, i.e., by displacing the acoustic rod. Using this method, the sound velocity was measured from 803K to 566K and the results are shown in Fig.1. Near the melting point it was found to be 1651±3 m/s, in agreement with the previously measured value of 1650±4 m/s.[23,32] It was also found that the sound velocity decreases monotonously with a non-linear temperature dependence, similar to previous studies.[20,21,23,27,33] Relative measurement of the sound velocity was carried out from slightly above the melting temperature and into the supercooled region in the temperature range from 566K to 520K. The measurement was terminated due to solidification of the liquid. The results of these measurements are presented in Fig. 1. It can be seen that the plateau in the sound velocity (1651±3 m/s) near the melting point begins approximately at 580K and remains constant within the experimental uncertainty to the minimum temperature achieved of 520K. Within the experimental uncertainty the temperature coefficient of the sound velocity is 0.00±0.05 m/s K, at least one order of magnitude less than its average value obtained at higher temperatures.[23,31]

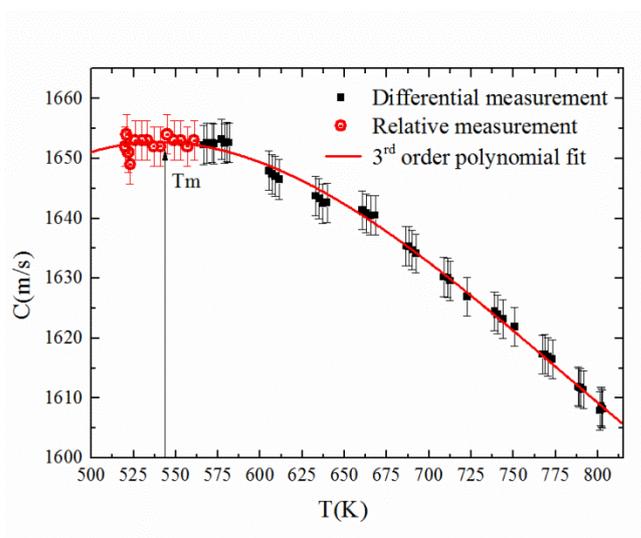

**FIG. 1.** Sound velocity as a function of temperature in liquid Bi, the melting temperature is indicted by Tm. Differential measurements taken in the normal liquid

state (squares) and relative measurements in the supercooled range (circles) as explained in the text. Line is 3$^{rd}$ order polynomial fit to the data, indicating the possibility of a broad maximum in the sound velocity.

## B. Neutron diffraction from liquid Bi

Static structure factors, *S(q)*, of liquid Bi were obtained from neutron diffraction measurements upon cooling the sample in the temperature range from 653K to 528K and are presented in Fig. 2. The lower bound of the temperature range was determined by the spontaneous solidification of the supercooled liquid and is slightly higher than the lowest temperature obtained in the ultrasonic sound velocity measurements, mainly due to the different sample size used in the two experiments. Radial distribution functions, *g(r)*, derived from these data by the Fourier transform methodology presented in Ref. 28, are presented in Fig. 3. The typical uncertainty in *g(r)* obtained by this method is of the order of a few percent.[28] It can be seen that as the liquid is cooled through the melting point and into the supercooled region, there is no discontinuous change is observed in the liquid structure, or any other major change which may be associated with rearrangement of the liquid. Instead, the liquid structure is modified continuously and smoothly, as can be seen in the vicinity of the shoulder on the r.h.s. of the main peak which grows as the temperature is decreased. This effect is illustrated in Fig. 4 which depicts the derivative of the RDF in the shoulder region. In the latter, we can observe a continuous increase in the value of *g'(r)* at 4.4Å as the shoulder grows in magnitude similar to Sb, but not quite achieving a sub peak which would require *g'(r)=0*.

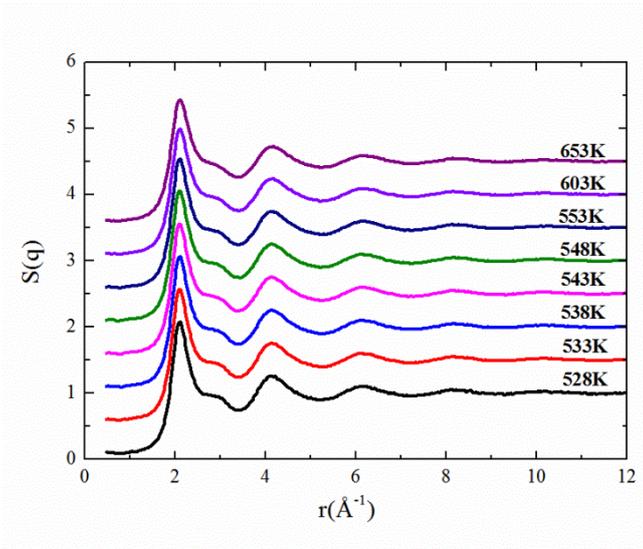

**FIG. 2.** The static structure factor of liquid Bi measured at selected temperatures during cooling into the supercooled region. Structure factors of successive temperatures are shifted by 0.5 units.

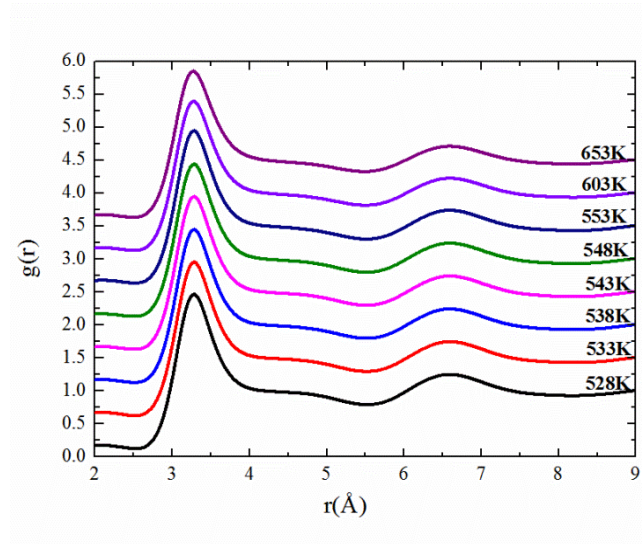

**FIG. 3.** The radial distribution functions of liquid Bi obtained from neutron diffraction data using the methodology described in Ref. 28. Successive curves are shifted by 0.5 units.

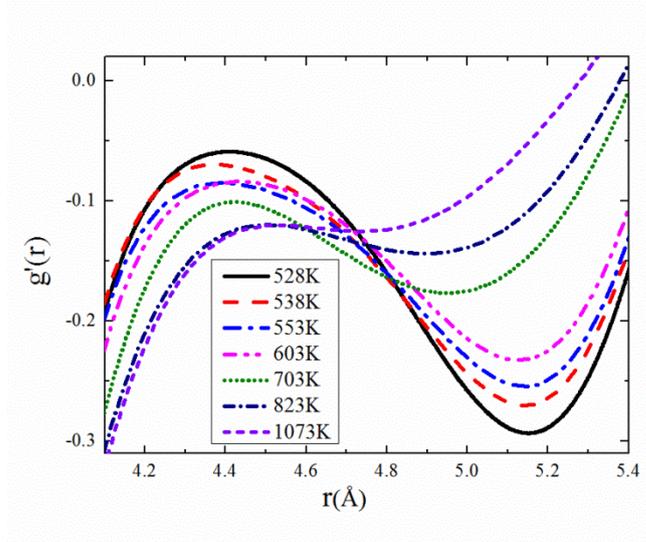

**FIG. 4.** Enlargement in the shoulder region of the slope of the first derivative of the radial distribution function $g'(r)$ of liquid Bi as a function of temperature (data for higher temperatures was taken from Ref. 30). The slope varies with distance, first increasing, then decreasing and finally increasing again, which is a signature of the shoulder. As the temperature decreases the maximum in the slope approaches zero from below indicating an evolution from a smooth radial distribution function, through a shoulder towards the formation of an additional sub-peak at lower temperatures.

### C. QCM analysis of g(r)

The quasi-crystalline model (QCM) for a liquid[30,34,35,36,37] is based on the realization that the radial distribution function, $g(r)$, of a solid crystal, may be modeled by summing the contributions of spherical shells of atoms with, the $i_{th}$ shell located at a distance $r_i$ from a reference atom located at the origin. At finite temperatures and still in an ideal crystal, each such contribution is smeared out in a Gaussian form with variance $\sigma_i$ as may be expected for solids at high temperature. In a solid crystal, the variance $\sigma_i$, originates from thermal excitations, is determined by the temperature and the elastic restoring force. Close to the melting temperature, the variance is found to increase with radial separation even in the solid phase.[38] The QCM of a liquid structure assumes that the RDF of the liquid may be obtained in terms of an underlying crystalline lattice, by allowing the Gaussian distribution width to increase with increasing distance from a reference atom at the origin. Thus, at large distances, the structure is smeared out, namely uncorrelated and there remains no indication of the underlying crystalline order. Different derivations of the QCM model exist in the literature,[34,35,36,37] leading in principle to the same functional form:

$$g(r) = \sum_{i=1}^{\infty} \frac{n_i V}{4\pi r_i^2} \frac{1}{\sqrt{2\pi \sigma_i^2}} \exp\left(-\frac{(r-r_i)^2}{2\sigma_i^2}\right), \qquad (1)$$

where $V$ is the atomic volume and $n_i$ is the number of atoms on the $i_{th}$ shell of atoms, located at radius $r_i$ about the origin with variance $\sigma_i$.

In previous work,[30,39] it has been shown that in order to represent the shape of the experimental radial distribution function of some liquid pnictides, the QCM must be modified to allow the Gaussian width of the first few shells to vary independently, whereas further variances may vary linearly as [36,40]:

$$\sigma_i^2 = r_i \cdot \beta^2 + \alpha. \qquad (2)$$

The QCM model defined in Eq. (1) is fitted to the experimentally observed $g(r)$'s using a least squares procedure to determine the best values of the fitting parameters. The A7 structure, a rhombohedral lattice with a basis of two atoms, is used as the underlying lattice to investigate the SRO in elements of column V.[39] Two parameters determine the A7 structure, $\alpha$ and $u$. The primitive vectors are given by $\vec{a_1} =$ $a(\epsilon, 1,1)$, $\vec{a_2} = a(1, \epsilon, 1)$ and $\vec{a_3} = a(1,1, \epsilon)$; where the relation between $\epsilon$ and the shear angle, $\alpha$, is given by $\epsilon = [1 - (1 + \cos \alpha - 2cos^2\alpha)^{1/2}]/\cos \alpha$. The two basis atoms are located at $\vec{d} = \pm u(\vec{a_1} + \vec{a_2} + \vec{a_3})$. It is noted that a simple cubic structure is given by $\alpha = 60^o$ and $u = 0.25$, a diamond cubic structure is given by $\alpha = 60^o$ and $u = 0.125$[41] and the low-temperature (4K) structure of As has $\alpha = 54.55^o$ and $u = 0.2276$.[42] Following Ref. 30, we represent the A7 structure by two independent subshells, where two free parameters, $\gamma_1$ and $\gamma_2$, are introduced into the model such that the positions of the first two Gaussians are given by $r_1' = r_1 - \gamma_1$ and $r_2' = r_1 + \gamma_2$. All subsequent shells from the third onwards are represented in our model by an ideal A7 structure i.e, u=0.25. Following the procedure of Ref. 30, the simple linear model relating the variances, e.g. Eq. (2), was modified and the variances of the first 8 shells, $\sigma_{1a}$, $\sigma_{1b}, \sigma_2, \sigma_3, \sigma_4, \sigma_5, \sigma_6$ and $\sigma_7$ were allowed to vary independently of each other and linearly beyond (Eq. (2)). The results of the fit to liquid Bi RDF are represented in Table I. The uncertainty of the fitting parameters was estimated in Ref. 30 to be 10% in the splitting of the two first subshells, 5% in the variances ($\sigma$'s) and 1% in the lattice vectors (a).

**TABLE I**. Values of the fitting parameters for liquid Bi, obtained from a least square fit of the modified QCM to the radial distribution functions.

| T[K] | 528 | 533 | 538 | 543 | 548 | 553 | 603 | 653 |
|---|---|---|---|---|---|---|---|---|
| $\gamma_1$[Å] | -0.28 | -0.27 | -0.27 | -0.32 | -0.32 | -0.32 | -0.34 | -0.36 |
| $\gamma_2$[Å] | 0.07 | 0.07 | 0.07 | 0.10 | 0.10 | 0.10 | 0.11 | 0.12 |
| $\sigma_{1a}$[Å] | 0.19 | 0.20 | 0.20 | 0.19 | 0.19 | 0.18 | 0.18 | 0.18 |
| $\sigma_{1b}$[Å] | 0.24 | 0.24 | 0.25 | 0.22 | 0.22 | 0.22 | 0.22 | 0.22 |
| $\sigma_2[\text{Å}]$ | 0.45 | 0.45 | 0.45 | 0.42 | 0.42 | 0.43 | 0.42 | 0.41 |
| $\sigma_3[\text{Å}]$ | 0.45 | 0.45 | 0.46 | 0.53 | 0.53 | 0.52 | 0.52 | 0.52 |
| $\sigma_4[\text{Å}]$ | 0.86 | 0.85 | 0.84 | 0.54 | 0.55 | 0.57 | 0.55 | 0.55 |
| $\sigma_5[\text{Å}]$ | 0.70 | 0.71 | 0.71 | 0.72 | 0.72 | 0.70 | 0.71 | 0.72 |
| $\sigma_6[\text{Å}]$ | 0.61 | 0.61 | 0.61 | 0.54 | 0.54 | 0.54 | 0.56 | 0.58 |
| $\sigma_7[\text{Å}]$ | 0.41 | 0.41 | 0.41 | 0.45 | 0.46 | 0.46 | 0.48 | 0.51 |
| $\alpha[\text{Å}^2]$ | -0.13 | -0.15 | -0.16 | -0.47 | -0.49 | -0.43 | -0.55 | -0.64 |
| $\beta[\text{Å}^{1/2}]$ | 0.35 | 0.35 | 0.35 | 0.37 | 0.38 | 0.37 | 0.39 | 0.41 |
| $a[\text{Å}]$ | 3.13 | 3.13 | 3.13 | 3.08 | 3.08 | 3.09 | 3.07 | 3.05 |

The radial distribution functions obtained from measurements of liquid Bi into the supercooled region were analyzed within the QCM. From the results presented in Table I, it is seen that there is no support for any structural rearrangement of the liquid within the temperature range in which the sound velocity remains constant. Instead, the change of the structure with temperature is both continuous and small. This conclusion is further strengthened by extending the temperature range considered well above the melting point based on earlier results reported in Ref. 8. Focusing only on the position of the first three shells, plotted in Fig. 5, we note that the trends in the temperature dependence of the radial shell positions remain unchanged over a very wide temperature range. We also note that the interatomic distance in the supercooled liquid continues to increase with decreasing temperature as previously observed in the melt[30] and in other liquid metals.[43]

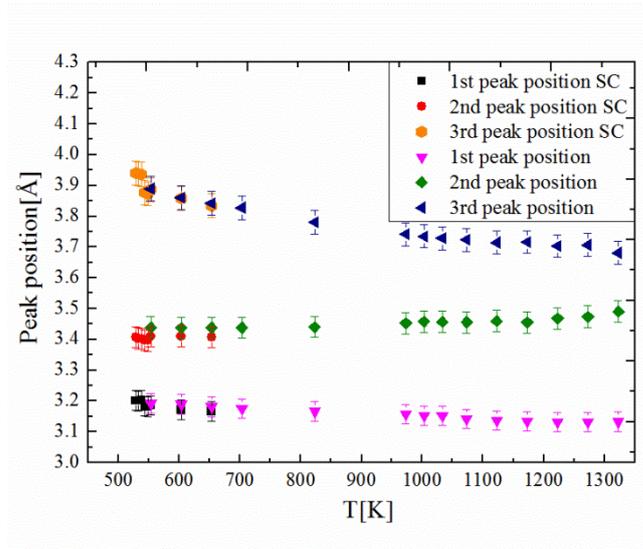

**FIG. 5.** Positions of the first three shells in the QCM model of the structure of liquid Bi as a function of temperature combining the present data (denoted SC) and that of Ref. 8. The supercooled peak positions are calculated from the fitted parameters in Table I.

### D. Liquid pnictide hypotheses

High accuracy sound velocity measurements as a function of temperature have been carried out in liquid Bi from the normal liquid state to the supercooled region. These measurements were obtained using the ultrasonic pulse echo technique. Comparison of our sound velocity results with previous studies[23] finds them to be in good agreement. The sound velocity remains at 1651±3 m/s constant within the experimental uncertainty from approximately 583K for a range over 60 degrees upon cooling. Previous studies have suggested that this plateau is associated with a structural rearrangement in the liquid state.[20,24,25,26] We suggest that this phenomenology reflects a broad maximum in the sound velocity, analogous to that observed in liquid Sb and is typical to column V in the periodic table. To estimate the behavior of the sound velocity near the melting point, we have fitted the temperature dependence to a third-order polynomial (Fig. 1). Third-order was chosen to reflect the expected asymmetry about a sound velocity maximum as observed in liquid Sb.[10,33] From this fit we find that the maximum in the sound velocity is at approximately 534K, very close to the minimum temperature achieved in our experiments. Further support for the possible existence of a broad maximum is provided by the results of sound velocity measurements in Bi-Sb alloys, which reveal a broad maximum in all compositions considered.[31]

The structure factor of liquid Bi was determined in the range 528-653K. The data are of very good quality and agree very well with previous measurements.[8] However, there is no indication of any abrupt or significant continuous change which might support the structural rearrangement hypothesis. The radial distribution functions obtained from the measured structure factors exhibit only modest, continuous change in the vicinity of the shoulder on the r.h.s. of the main peak which increases with decreasing temperature. These changes in the liquid structure are, again, in agreement with similar changes observed in liquid Sb as a function of temperature near the sound velocity maximum.[27] Thus they support the liquid pnictide hypothesis that correlates sound velocity anomaly with the sub-structure of the radial distribution functions of group V elements. We have also calculated the coordination number as a function of temperature and the temperature dependence remains linear in the supercooled region as previously reported in Ref. 8. The modified QCM model[30] was applied to analyze the liquid Bi structure in the supercooled region and the results of the fit were of a similar quality to Ref. 30. It was found that the parameters of the model change continuously throughout the measured temperature range. Furthermore, extending the comparison to higher temperatures based on previously acquired data[8] and its analysis[30] shows that the trends of the temperature dependence remain unchanged, i.e., as the temperature decreases the splitting between the first two subshells in the liquid structure decreases, reflecting a weakening of the Peierls distortion as discussed in Ref. 30. In addition the position of the third shell shifts rapidly away from the origin, accounting for the buildup of the shoulder. In passing, we note that the two experiments and their subsequent QCM analysis are of sufficiently high quality that the results can be compared quantitatively. We conclude that there is no observed well defined structural rearrangement in liquid Bi near the melting point and that the plateau in the sound velocity probably reflects a broad maximum similar to that observed in other liquid pnictide systems. We therefore expect, that at lower temperatures, if were experimentally accessible, the sound velocity will decrease. This conclusion is in agreement with high pressure measurements[44] showing that at pressures of several hundred MPa the sound velocity of liquid Bi exhibits a maximum at approximately ambient melting temperature. Confirmation of this hypothesis will require further experimental studies of a new design.

## IV. CONCLUSIONS

The sound velocity in liquid Bi exhibits a plateau near the melting temperature, uniquely among the elements. By extending the sound velocity measurements for the first time into the supercooled region, we have found that this plateau extends over ca. 60 degrees, throughout the accessible temperature range into the supercooled state. We have performed structural studies by neutron diffraction in the supercooled state and analyzed them within the QCM. This analysis does not indicate any non-continuous significant structural rearrangement in the liquid, as proposed by several previous studies, but rather a continuation of changes in the liquid structure observed in the melt. Thus, we conclude that the observed plateau in the sound velocity is part of a broad maximum in the sound velocity typical of liquid pnictides and that no structural rearrangement takes place in the vicinity of the melting temperature of liquid Bi, neither above nor below it.

**ACKNOWLEDMENTS**

The authors wish to acknowledge the allocation of beam time at Laboratoire Léon Brillouin CEA/Saclay, and the support of CHE-IAEC research grant by the Pazy foundation.